# Entanglement-free Metrology Exploiting Multimode Hong–Ou–Mandel Sensor Advantage

Qian Li and Jianning Han

School of Physics and Electronic Engineering,

Shanxi Normal University, Taiyuan 030031, China

E-mail: 20250085@sxnu.edu.cn; E-mail: hanjn46@nuc.edu.cn

**Abstract:** The Hong-Ou-Mandel (HOM) interference in the multimode frequency domain has been explored for precision metrology, with several experimental demonstrations exploiting its robustness against dispersion and phase noise, as well as its large dynamic range and compatibility with fragile samples. Conventional multimode HOM metrology exploits frequency-entangled states, which naturally satisfy bosonic exchange symmetry under any centered symmetric joint spectral distribution, to provide these advantages. However, these entangled states are typically generated via spontaneous parametric down-conversion (SPDC), requiring strong pump lasers that hinder practical implementation. In this paper, we employ frequency product states, which do not possess entanglement or path-mode exchange symmetry, as the probe state and post-select measurement outcomes exhibiting frequency anti-correlation. Our results demonstrate that these advantages—peak narrowing, dispersion cancellation, phase-noise immunity, a large dynamic range, and compatibility with fragile samples—arise neither from entanglement nor from bosonic exchange symmetry, but rather from spectral anti-correlation. We further show that entanglement is not the source of the measurement precision: the entanglement-free approach attains the same quantum Fisher information as the entangled-state scheme, indicating that the fundamental precision limit does not originate from entanglement.

## 1 Introduction

The HOM interference, first demonstrated in 1987 [1], has evolved into a cornerstone of quantum photonics. Its fundamental principle, whereby two indistinguishable photons incident on a beam splitter bunch at the output ports, has found extensive applications in quantum computing [2-4], quantum state verification [5,6], quantum communication [7,8], and has shown promise in quantum metrology [9,10]. In the context of sensing and imaging, the multimode HOM interferometer has attracted considerable interest because it offers several metrological advantages that are potentially useful in practical applications. These include robustness against even-order dispersion [11,12], immunity to phase noise [13,14], a broad dynamic measurement range that circumvents the $2\pi$ phase ambiguity [1,15], and a twofold enhancement in axial resolution [16,17]. These features—robustness against dispersion, immunity to phase noise, a broad dynamic range, and enhanced axial resolution—have motivated a range of experimental demonstrations, including time-delay measurement [18-20], polarization sensing [21-23], radar detection [24,25], quantum microscopy [26,27], and quantum optical coherence tomography (QOCT) [16,28] and so on. Conventionally, these metrological advantages are attributed to the use of spectrally entangled biphoton states, typically generated via spontaneous parametric down-conversion (SPDC). In such states, the frequency of one photon is perfectly correlated with that of the other, a property that naturally gives rise to the bosonic exchange symmetry crucial for the HOM effect [29-31]. However, SPDC is an inherently inefficient probabilistic process that requires high-power pump lasers and nonlinear crystals, posing a significant barrier to practical, robust, and cost-effective implementations. This reliance on a complex quantum resource has led to the widespread belief that entanglement is the fundamental prerequisite for achieving the aforementioned metrological advantages.

This paradigm has been challenged by a series of classical emulations. Notably, Kaltenbaek et al. demonstrated HOM-like interferometry using chirped laser pulses [32], inspiring further efforts to reproduce dispersion cancellation and resolution enhancement using classical light sources [33-40]. These intriguing results prompted Franson to argue, on qualitative grounds, that HOM interference is a local process that admits classical simulation [41]. This interpretation has been adopted by subsequent studies [42-44]. Nevertheless, a quantitative theoretical framework that explains why these advantages arise has remained elusive. Most existing studies treat classical emulations as isolated phenomena within a purely classical framework, without providing a direct quantitative proof or explanation from a quantum-mechanical perspective that connects them to the entanglement-based case.

In this paper, we establish such a quantitative framework by demonstrating that all the metrological advantages of HOM interferometry—dispersion cancellation, phase-noise immunity, peak narrowing, and wide dynamic range—can be reproduced without entanglement. We replace the spectrally entangled biphoton state with a frequency product state, a simple, entanglement-free state. By applying post-selection to extract anti-diagonal components from a spectral correlation measurement, we show that the resulting interference pattern is mathematically identical to that obtained with entangled states. Our analysis rigorously proves that the essential resource for dispersion cancellation, phase-noise immunity and peak narrowing is not entanglement itself, but effective frequency anti-correlation, which can be synthesized from product states via post-selection. We further show that the measurement precision does not originate from entanglement either, the entanglement-free approach achieves the same quantum Fisher information as the entangled-state scheme. This provides, for the first time, a rigorous mathematical counterpart to Franson's qualitative argument and fully bridges the gap between entangled HOM interferometry and its classical emulations. Importantly, our proposed method is not merely a theoretical abstraction; its practical viability has been demonstrated in a specific application. In a parallel effort, we have successfully applied this post-selection scheme to quantum optical coherence tomography (QOCT) using a weak multimode coherent state [45]. The present work, however, goes significantly beyond that specific application. While the QOCT demonstration serves as a concrete proof of concept, our aim here is to reveal the universal physical mechanism underlying all HOM-based sensing. We generalize the scheme to a canonical HOM interferometer model and show that the post-selection-induced spectral anti-correlation is a general principle that applies to any HOM-type measurement, regardless of the specific imaging modality.

The structure of this paper is as follows. In Section 2, we present the general scheme of the product-state HOM interferometer with post-selection and derive the interference pattern, which we show to be identical to that of the entanglement-based case. In Section 3, we systematically demonstrate that our scheme preserves the key metrological advantages, peak narrowing and wide-range measurement (3.1), robustness against dispersive samples (3.2), and immunity to phase noise (3.3). In Section 3.4, we analyze the Fisher information of the scheme and clarify its precision cost relative to the entangled-state HOM interferometer. In Section 4, we conclude with a discussion on the implications of our findings for quantum metrology.

## 2 The Scheme of HOM Interferometer

As shown in Fig. 1a, the conventional measurement technique based on HOM interference primarily employs the spectrally entangled biphoton state generated through the interaction between a strong pump light and a nonlinear crystal, which can be expressed as,

$$|\psi\rangle_{HOM} = \int d\Omega \varphi(\Omega) \hat{a}^{+}(\omega_0 + \Omega) \hat{b}^{+}(\omega_0 - \Omega) |0\rangle \tag{1}$$

where $\varphi(\Omega)$ represents the spectral distribution function of the entangled biphoton state. The operator $\hat{a}^{+}(\omega_0 + \Omega)$ denotes the creation of a photon with frequency $\omega_0 + \Omega$ in path mode $a$, while $\hat{b}^{+}(\omega_0 - \Omega)$ denotes the creation of a photon with frequency $\omega_0 - \Omega$ in path mode $b$. When the frequency of one photon is determined, the frequency of the other photon is simultaneously determined. Then, the photon of mode $a$ enters the beamsplitter (BS) in Fig. 1a after a time delay of $\tau_c$, while the photon of mode $b$ interacts with the sample. The phase that carried by the photon of mode $b$ after interaction can be given by $\exp(i(\omega_0 - \Omega)\tau)$, where $\tau$ is the time delay introduced by interaction between photon and the sample. Choosing a Gaussian spectral function $|\varphi(\Omega)|^2 = \exp\left(-\Omega^2/(2\sigma^2)\right)$, the result after the coincidence measurement can be expressed as,

$$P_{HOM} = \frac{1}{2} - \frac{1}{2} e^{\left(-2\sigma^2(\tau - \tau_c)^2\right)}, \tag{2}$$

The detailed derivation is provided in Supplemental Document A. As $\tau$ gradually approaches $\tau_c$, the coincidence probability $P_{HOM}$ gradually decreases. When the time delay difference between the two arms becomes zero, that is, when $\tau = \tau_c$, the coincidence probability reaches its minimum value of zero, which is the well known HOM dip phenomenon. The position of the HOM dip corresponds one to one with the magnitude of the unknown phase to be measured. This forms the basis for using HOM interference to measure or image unknown samples.

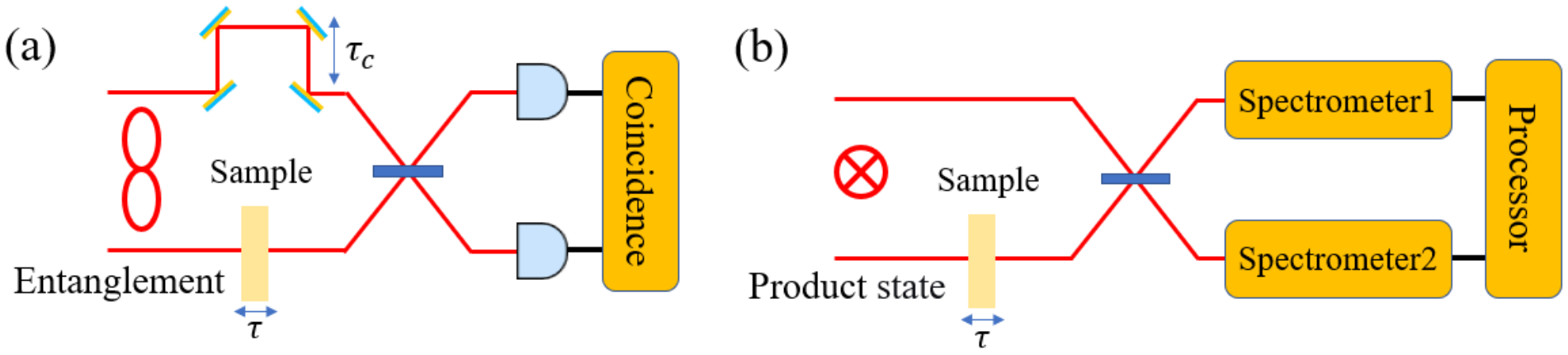

Fig. 1**.** Schematic of the HOM interferometer with (a) an entangled biphoton state and (b) a frequency product state. (a) A frequency-entangled photon pair is generated by the entanglement source. One photon is delayed by an adjustable time $\tau_c$, while the other interacts with the sample, which introduces an unknown time delay $\tau$. The two photons interfere at the beam splitter (BS), and the coincidence measurement records the two-photon coincidence events at the two output ports. (b) A frequency product state, in which the two photons are independent in frequency, is used as the probe. One photon passes through the sample, and the two photons interfere at the BS. The two output ports are connected to Spectrometer 1 and Spectrometer 2, which measure the frequencies $\omega$ and $\omega'$ of the two photons, respectively. The processor then performs post-selection by extracting the anti-diagonal components from the spectral correlation measurement.

Then, we demonstrate that even if the probe state is not the entangled state shown in Eq. (1), a measurement result similar to that in Eq. (2) can still be obtained. Specifically, as illustrated in Fig. 1(b), we employ a frequency product state as the probe state, which can be expressed as,

$$|\psi\rangle_{in} = \int d\omega_1 \varphi_p(\omega_1)\hat{a}^+(\omega_1)|0\rangle \otimes \int d\omega_2 \phi_p(\omega_2)\hat{b}^+(\omega_2)|0\rangle, \tag{3}$$

where $\varphi_p(\omega_1)$ and $\phi_p(\omega_2)$ represent the spectral distribution function of the photons in mode $a$ and mode $b$, respectively. It can be seen that the photons in mode $a$ and mode $b$ are independent of each other in the frequency degree of freedom, meaning that there is no frequency entanglement between the two photons. Consequently, the frequency of the photon in mode $b$ cannot be determined by measuring the frequency of the photon in mode $a$. Similar to Fig. 1(a), we place the sample to be measured in the path of the mode $b$ photon. The phase acquired by the photon in mode $b$ after the interaction can be expressed as $\exp(i\omega_2\tau)$. The HOM interference directly measures this time delay $\tau$. For a sample of thickness $l_s$ and refractive index $n$, the delay is related to the thickness by $\tau = nl_s/c$. Thus, $\tau$ is the fundamental measured quantity, and the sample thickness $l_s$ can be retrieved from $\tau$ once the refractive index $n$ is known. Then, the photon in mode $a$ overlaps with the photon in mode $b$ at the BS, and the output state can be expressed as,

$$|\psi\rangle_{out} = \frac{1}{2}\int d\omega_1 d\omega_2 \varphi_p(\omega_1)\phi_p(\omega_2)e^{i\omega_2\tau}\big(c^+(\omega_1) + d^+(\omega_1)\big)\big(c^+(\omega_2) - d^+(\omega_2)\big)|0\rangle \tag{4}$$

The operators $\hat{c}^+(\omega)$ and $\hat{d}^+(\omega)$ denote the creation operators for the two output ports ($c$ and $d$) of the BS, respectively. The frequencies $\omega_1$ and $\omega_2$ refer to the optical fields associated with paths $a$ and $b$, respectively.

Finally, spectral correlation measurements are performed on the output optical field described by Eq. (4). This measurement technique can be implemented using several established methods. As an example, a scanning spectrometer may be employed to convert frequency information into spatial information, after which the intensity at different positions is recorded [46,47]. In this context, the spectral correlation measurement operator can be expressed as,

$$\hat{P} = |1(\omega)\rangle_{c\,c}\langle 1(\omega)|\otimes|1(\omega')\rangle_{d\,d}\langle 1(\omega')|, \tag{5}$$

which represents the joint projection operator that measures a photon with frequency $\omega$ in path mode $c$ and a photon with frequency $\omega'$ in path mode $d$. In Fig. 1(b), the measurement operator corresponding to Eq. (5) is implemented using two spectrometers, denoted as Spectrometer 1 and Spectrometer 2. It should be noted that in the experiment, the measurement operator corresponding to Eq. (5) does not actually require the photons in path mode $c$ and path mode $d$ to arrive at the two spectrometers strictly simultaneously. It suffices that both photons arrive at the spectrometers within a certain time window. Therefore, even when the input state is the frequency product state given by Eq. (4), the corresponding measurement result can still be obtained as long as the time window is properly set. This result can be expressed as,

$$P_c(\omega,\omega') = \langle\psi|\hat{P}|\psi\rangle = \frac{1}{4}\left|\varphi_p(\omega')\phi_p(\omega)e^{i\omega\tau} - \varphi_p(\omega)\phi_p(\omega')e^{i\omega'\tau}\right|^2, \tag{6}$$

It can be found that $P_c$ is the function of $\omega$ and $\omega'$. To provide a clearer description of the results corresponding to Eq. 6, we present the simulation results of $P_c$ in Fig. 2a. In the simulations, the light source has a central wavelength of 810 nm, and both spectral distributions $\varphi_p(\omega)$ and $\phi_p(\omega)$ are modeled as Gaussian functions with a full width at half maximum (FWHM) of 50 nm. The sample thickness $l_s$ is chosen to be 100 μm with a refractive index $n = 1$, which corresponds to a time delay of $\tau = nl_s/c = 1/3ps$. Here, $n = 1$ corresponds to air/vacuum and is used as a simplified reference case. In the non-dispersive case, the refractive index only sets the relation $\tau = nl_s/c$ and does not affect the shape, width, or precision of the final interference pattern. $l_s$ is the geometric thickness of the sample, and $\tau$ is the time delay directly measured by the HOM interferometer. The two quantities are related by $\tau = nl_s/c$. The thickness is selected so that the induced delay is larger than the temporal width of the HOM interference peak, i.e., the wave-packet FWHM, ensuring that the interference peak can be clearly resolved. Retrieving $l_s$ from the measured $\tau$ is important because the thickness is a fundamental geometric property of the sample, and its measurement enables depth-resolved imaging, ranging, and material characterization. Using these parameters, we obtain simulation results based on Eq. (6), which depend on both $\omega$ and $\omega'$.

Then, we select the components of the measurement result $P_c(\omega,\omega')$ that simultaneously satisfy $\omega = \omega_0 + \Omega$ and $\omega' = \omega_0 - \Omega$. This selection process, performed in the Processor section shown in Fig. 1(b), corresponds to retrieving data along the anti diagonal line from the spectral correlation measurement results associated with Eq. (6). In the Fig. 2a, the yellow dashed line represents this operation. In fact, taking anti diagonal elements from spectral correlation measurements effectively mimics the behavior of photon pair spectral entanglement. The post selected result originates from the interference of photon pairs that satisfy the correlation relation $\omega + \omega' = 2\omega_0$. The output corresponding to the anti-diagonal elements from spectral correlation measurements can be expressed as,

$$P_{anti} = \frac{1}{2} e^{-\frac{\Omega^2}{(2\sigma^2)}} (1 - \cos(2\Omega\tau)), \quad (7)$$

In the calculation process, both $\varphi_p(\omega)$ and $\phi_p(\omega')$ are set to identical Gaussian distributions, given by $\varphi_p(\omega) = \exp(-(\omega - \omega_0)^2/(8\sigma^2))$, $\phi_p(\omega') = \exp(-(\omega' - \omega_0)^2/(8\sigma^2))$. The detailed derivation is provided in Supplemental Document B. The simulation results corresponding to Eq. 7 are presented in Fig. 2b. In Fig. 2b, the horizontal axis shows how much the photon frequency deviates from the central frequency $\omega_0$, and the vertical axis gives the normalized intensity of the optical field at each corresponding frequency. Finally, we can obtain the final output result by performing the Fourier transform on the results derived from Eq. 7, which can be expressed as,

$$P \propto \frac{1}{2} e^{-\frac{\sigma^2 t^2}{2}} - \frac{1}{4} e^{-\frac{\sigma^2 (t-2\tau)^2}{2}}, \quad (8)$$

During the Fourier transform of Eq. 7, only the positive frequency component of the result is retained. From Eq. 8, two peaks are identified at positions $t = 0$ and $t = 2\tau$. The FWHM of both peaks is $2\sqrt{2ln2}/\sigma$. In contrast, for the entangled HOM peak obtained from Eq. (2), the FWHM is $\sqrt{2ln2}/\sigma$. However, since the peak from Eq. 8 appears at $t = 2\tau$, a variable substitution $t = 2\tau'$ is applied to Eq. 8, yielding

$$P_{t=2\tau} = \frac{1}{2} e^{-\frac{2\sigma^2 \tau'^2}{2}} - \frac{1}{4} e^{-2\sigma^2 (\tau' - \tau)^2}, \quad (9)$$

Eq. 9 produces a second peak whose position and FWHM exactly match those of the HOM interference pattern in Eq. 2. From the position and amplitude of this peak, the sample thickness is determined. Fig. 2c presents the corresponding simulation results based on Eq. 8. For clarification, we have transformed the $x$-axis in Fig. 2c into spatial coordinates. It can be observed that a total of two peaks appear in Fig. 2c located at positions 0 μm, and 200 μm respectively—corresponding precisely to those obtained from Eq. 6 at $t = 0$ and $2/3$ ps.

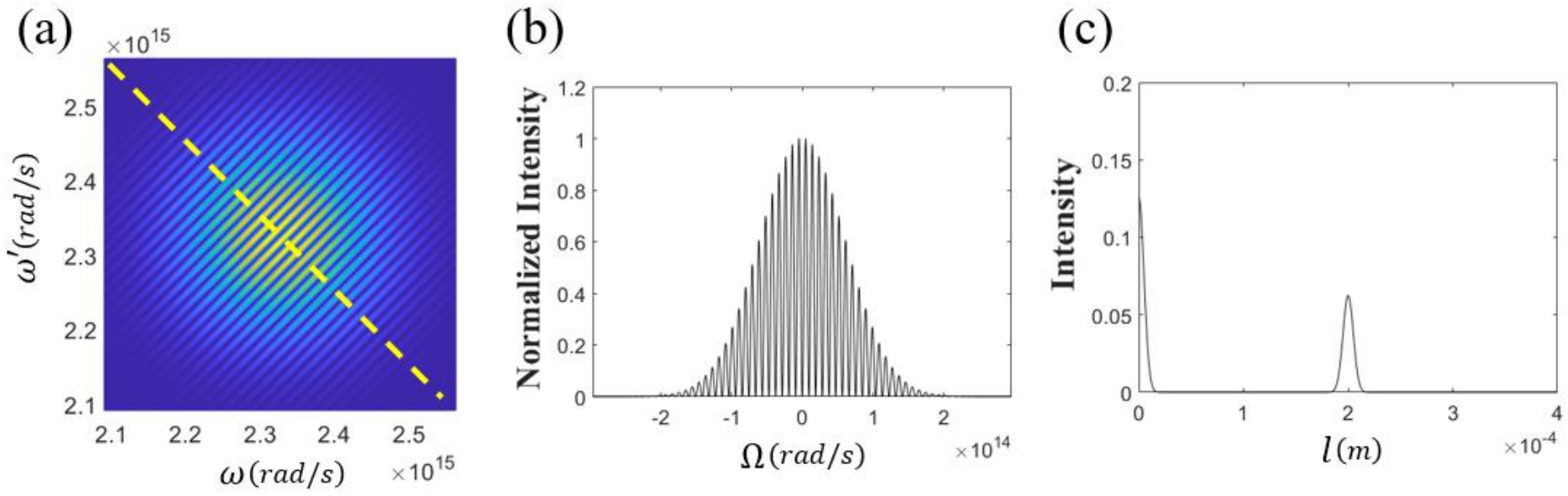

Fig. 2. Simulation results based on frequency product state and spectral correlation measurements. (a), Simulated spectral correlation map $P_c(\omega, \omega')$. The yellow dashed line indicates the anti-diagonal selection ($\omega = \omega_0 + \Omega$, $\omega' = \omega_0 - \Omega$). (b), Post-selected intensity along the anti-diagonal as a function of frequency detuning $\Omega$. (c), Fourier transform result of Fig. 2b in the spatial domain. Two peaks appear at 0 μm and 200 μm, where the second peak determines the sample thickness.

## 3 Metrological advantages of the product-state HOM interferometer

In the preceding section, we have demonstrated that employing frequency product states as the probe jointly with a post-selection measurement scheme yields interference peaks whose central positions and FWHM are identical to those obtained from entanglement-based HOM interference. This result establishes that the essential features of HOM interference can be faithfully reproduced without employing entangled states. Building upon this finding, in this section we further prove that the interference peaks generated by the proposed product-state plus post-selection scheme retain all the well-established metrological advantages of conventional HOM interferometry. Specifically, we demonstrate that the scheme exhibits robustness against dispersion, immunity to phase noise, halved FWHM corresponding to enhanced resolution, and a broad dynamic measurement range. These results provide strong evidence that the metrological superiority of HOM interference does not necessarily originate from quantum entanglement, but can be achieved using more readily accessible frequency product states combined with appropriate post-selection.

**3.1 Peak Narrowing and Wide-Range Measurement**

In measurement and imaging techniques based on HOM interference, a well recognized advantage is the enhancement of resolution. Specifically, the FWHM of the HOM interference peak can be compressed to half of that corresponding to the Fourier transform of the two photon spectrum. This interference peak narrowing effect provides a twofold improvement in axial resolution in HOM interference based QOCT [12,16,17]. Specifically, when deriving the HOM interference result based on the frequency entangled state (Eq. 2), we assume that the biphoton joint spectral distribution function $\varphi(\Omega)$ follows a Gaussian distribution, i.e., $|\varphi(\Omega)|^2 = \exp\left(-\Omega^2/(2\sigma^2)\right)$. After Fourier transformation, the corresponding temporal wave packet has a FWHM of $2\sqrt{2ln2}/\sigma$. However, as shown in Eq. 2, the FWHM of the HOM temporal interference result based on the frequency entangled state is $\sqrt{2ln2}/\sigma$, which is exactly half the Fourier transform width of $|\varphi(\Omega)|^2$. The essence of this peak narrowing effect lies in the fact that two photon interference doubles the positional parameter corresponding to the unknown phase from $\tau$ to $2\tau$, thereby compressing the interference peak width in the time domain.

According to Eq. 7, it can be demonstrated that the scheme based on the frequency product state with post selection also exhibits a similar peak narrowing advantage. As shown in Eq. 7, $P_{anti} = 1/2\exp\left(-\Omega^2/(2\sigma^2)\right)(1-\cos(2\Omega\tau))$, where the cosine term depends on $2\Omega\tau$ rather than $\Omega\tau$. This dependence indicates that the anti diagonal component extracted by post selection effectively mimics the frequency anti correlation characteristic of the entangled biphoton state, causing the time delay $\tau$ to be doubled to $2\tau$ in the interference term. After Fourier transformation, the resulting temporal interference peak has a FWHM of $\sqrt{2ln2}/\sigma$, which is identical to that of the entangled based HOM interference result and is compressed by a factor of two compared to the Fourier transform width of the input spectrum, $2\sqrt{2ln2}/\sigma$. Therefore, although our scheme employs an unentangled frequency product state as the input, the post selection operation still reproduces a peak narrowing effect comparable to that of entangled HOM interference. The same conclusion can be obtained by directly comparing Eq. 9 with Eq. 2. Under the same photon spectral distribution, the entangled state based HOM interference result (Eq. 2) and the result of the frequency product state with post selection scheme (Eq. 9) exhibit identical temporal Gaussian peak forms. The Gaussian peaks in both expressions have exactly the same width of $1/(2\sigma)$ (corresponding to an FWHM of $\sqrt{2ln2}/\sigma$) and the same delay dependence. This demonstrates that by simulating frequency anti correlation through post selection, our entanglement free scheme can precisely reproduce the measurement output of entangled HOM interference in the time domain.

On the other hand, wide range measurement represents another important advantage of measurement techniques based on HOM interference. Specifically, due to the use of a broadband light source, the measurement result of HOM interference depends only on the relative time delay between the two arms, without requiring absolute phase information [15,27,48]. This fundamentally avoids the $2\pi$ phase ambiguity problem compared to traditional phase sensitive interferometers such as the Mach Zehnder interferometer, thereby enabling unambiguous and large range measurement of sample thickness or time delay. However, it should be noted that this advantage is not unique to HOM interference but is a common feature shared by all interference techniques that utilize low coherence light sources, including white light interferometry and optical coherence tomography [49-51]. This advantage is preserved in our scheme. As shown in Eq. 9, the temporal interference peak obtained after post selection has exactly the same form as that of entangled HOM interference, with its peak position uniquely determined by the sample delay $\tau$ without involving any periodic phase term. Therefore, our scheme can also achieve the same wide dynamic range unambiguous measurement as HOM interference.

### 3.2 Dispersion Samples

In this part, we discuss the advantages of our scheme when applied to dispersive samples. The dispersion robustness of HOM interference ensures that its measurement results are unaffected by the broadening of dispersive wave packets, a feature that is particularly valuable for detection and imaging using broad-spectrum optical fields [11,52,53]. When considering the dispersion of the sample, the phase carried by the signal optical field along path $b$ in Fig. 1 can be expressed as,

$$H(\omega) = e^{i\beta(\omega)l_s}, \tag{10}$$

In this expression, $l_s$ is the sample thickness, and $\beta(\omega)$ is the propagation constant of the optical field inside the sample at frequency $\omega$. Expanding $\beta(\omega)$ as a Taylor series around the central frequency $\omega_0$ yields $\beta(\omega_0 + \Omega) = \omega_0 n(\omega_0)/\mathrm{c} + \beta_1\Omega + \beta_2\Omega^2$, where $\beta_1$ and $\beta_2$ are the first order and second order dispersion coefficients of the sample, respectively. Higher order dispersion effects are typically negligible in practical situations. This is particularly true for biological tissue imaging [16,54,55]. Replacing the sample in the entanglement-based HOM interferometer with the dispersive sample described by Eq. (10), the coincidence measurement result can be expressed as,

$$P_{S,HOM} \propto \frac{1}{2} - \frac{1}{2} e^{\left(-2\sigma^2(\beta_1 l_s - \tau_c)^2\right)}, \tag{11}$$

It can be observed that the interference peak appears at $\tau_c = \beta_1 l_s$, with a FWHM of $\sqrt{2ln2}/\sigma$. Notably, the peak remains unaffected by the second-order dispersion coefficient $\beta_2$ of the sample. This insensitivity arises from the presence of the interference term $H(\omega_0 - \Omega)H^*(\omega_0 + \Omega)$ and its complex conjugate in the two-photon interference result, which cancels out even-order dispersion effects. The detailed derivation is provided in Supplemental Document C.

Next, by replacing the sample in Fig. 1(b) with the dispersive sample described by Eq. (10), we obtain the spectral correlation measurement result for the case where the probe state is a frequency product state and the sample is dispersive. This result can be expressed as,

$$P_{S,c}(\omega,\omega') = \frac{1}{4}\left|\varphi_p(\omega')\phi_p(\omega)H(\omega) - \varphi_p(\omega)\phi_p(\omega')H(\omega')\right|^2, \tag{12}$$

We then perform the same post-selection operation as in part 2, i.e., we select only the anti-diagonal components satisfying $\omega = \omega_0 + \Omega$ and $\omega' = \omega_0 - \Omega$. Substituting these into Eq. (12) and assuming identical Gaussian spectral distributions $\varphi_p(\omega) = \exp\left(-(\omega - \omega_0)^2/(8\sigma^2)\right)$, $\phi_p(\omega') = \exp\left(-(\omega' - \omega_0)^2/(8\sigma^2)\right)$, we obtain

$$P_{S,anti} = \frac{1}{2}e^{-\frac{\Omega^2}{(2\sigma^2)}}\left(1 - Re(H(\omega_0 + \Omega)H^*(\omega_0 - \Omega))\right), \tag{13}$$

Crucially, the interference term $H(\omega_0 + \Omega)H^*(\omega_0 - \Omega)$ appearing in Eq. (13) is exactly the same as that in the entanglement-based HOM case. Using the Taylor expansion of $\beta(\omega)$, we have

$$H(\omega_0 + \Omega)H^*(\omega_0 - \Omega) \approx e^{2i\beta_1\Omega l_s}, \tag{14}$$

where the $\beta_2$ term cancels exactly. Substituting this into Eq. (13) and performing a Fourier transform, the final temporal interference pattern is

$$P_S \propto \frac{1}{2}e^{-\frac{\sigma^2 t^2}{2}} - \frac{1}{4}e^{-\frac{\sigma^2(t-2\beta_1 l_s)^2}{2}}, \tag{15}$$

Thus, our product-state scheme with post-selection reproduces the same dispersion-cancellation property as entanglement-based HOM interferometry. The post-selection operation effectively extracts the frequency-anti-correlated component from the product state, generating an effective interference term $H(\omega_0 + \Omega)H^*(\omega_0 - \Omega)$ that cancels the even-order dispersion. This demonstrates that the dispersion robustness of HOM-based metrology does not rely on entanglement but can be achieved using more accessible frequency product states combined with appropriate post-selection. Furthermore, we present the corresponding simulation results in Fig. 3. Fig. 3a shows the simulation results of classical first order low coherence interferometry under the influence of dispersion. The light source has a central wavelength of 810 nm and a spectral FWHM of 100 nm. The dispersion coefficients for the samples are $\beta_1 = 5 \times 10^{-9}\mathrm{sm}^{-1}$ and $\beta_2 = 88 \times 10^{-25}\mathrm{s}^2\mathrm{m}^{-1}$. For comparison, heavy flint glass typically exhibits a group-velocity dispersion in the range of $(1.0\sim 2.0) \times 10^{-25}s^2\, m^{-1}$ at 810 nm (e.g., SF10 has approximately $1.55 \times 10^{-25}s^2\, m^{-1}$) [16,56]. The value used in our simulation is therefore about $44\sim 88$ times larger than that of heavy flint glass, representing a strongly dispersive scenario sufficient to demonstrate the dispersion-robustness of our scheme. In Fig. 3a, $l$ is the spatial coordinate obtained from the measured time delay through $l = \tau/\beta_1$. The three curves correspond to sample thicknesses of $2 \times 10^{-5}m$, $4 \times 10^{-5}\, m$, and $6 \times 10^{-5}\, m$, respectively. These values are chosen as typical thicknesses relevant to biomedical imaging and industrial inspection. They correspond to time delays $\tau = \beta_1 l_s$ that are larger than the wave-packet FWHM, so that the interference peaks are clearly resolvable. The simulation results reveal peaks appearing at the corresponding positions, but the peak width broadens as the sample thickness increases due to the effect of sample dispersion. Similarly, Fig. 3(b) presents the simulation results of our product state with post selection scheme under the same sample parameters used in the classical first order low coherence interferometry. Notably, the FWHM of the peaks at the corresponding positions remains unaffected by sample dispersion, which is identical to the dispersion robustness exhibited by entangled HOM interference.

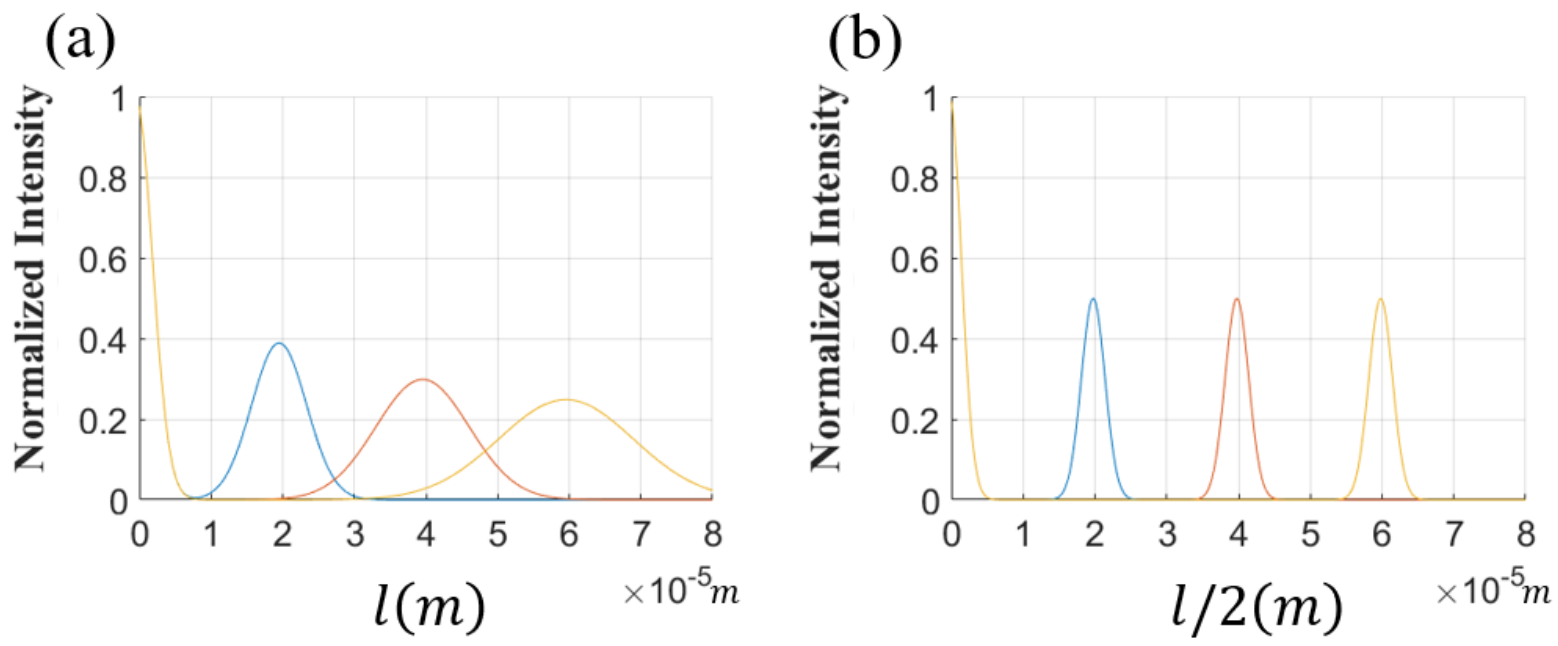

Fig. 3, (a). Simulation results of classical first order low coherence interferometry. The peak width broadens as the sample thickness increases due to dispersion. (b). Simulation results of the product state with post selection scheme. The peak width remains unaffected by sample dispersion, exhibiting the same dispersion robustness as entangled HOM interference.

### 3.3 phase perturbations

In addition to dispersion robustness, another well-recognized advantage of HOM-based interferometry is its inherent immunity to phase noise [10,14]. In conventional phase-sensitive interferometers, such as the Mach–Zehnder interferometer, random phase fluctuations in the optical paths can significantly degrade measurement precision. In contrast, HOM interference relies on two-photon coincidence counting, and its interference term depends only on the relative time delay rather than the absolute optical phase [13]. This section demonstrates that our product-state scheme with post-selection preserves this advantage.

We consider random phase perturbations introduced by the sample or the environment. The phase noise can be modeled as a frequency-dependent phase shift $\exp(i\phi(\omega))$. To first order, it can be expanded around the central frequency $\omega_0$ as

$$e^{i\phi(\omega)} \approx e^{i\left(\phi(\omega_0)+\phi'(\omega_0)(\omega-\omega_0)\right)}, \tag{16}$$

In the two-arm configuration shown in Fig. 1, the phase noises accumulated in path $a$ and path $b$ are generally independent. For the entanglement-based HOM interferometer, following the same derivation as in Supplemental Document A but including the phase noise terms, the coincidence measurement result becomes

$$P_{P,HOM} = \frac{1}{2}\left(1 - e^{-2\sigma^2\left((\tau-\varepsilon_0)-\tau_c\right)^2}\right), \tag{17}$$

where $\varepsilon_0 = \phi_a'(\omega_0) - \phi_b'(\omega_0)$ represents the relative linear phase noise coefficient between the two paths. The detailed derivation is provided in Supplemental Document D. Eq. 17 shows that the effect of linear phase noise is simply to shift the HOM dip position by $\varepsilon_0$. Importantly, the shape and width of the dip remain unchanged. Since $\varepsilon_0$ represents a constant time offset, it can be calibrated out or compensated by adjusting $\tau_c$. Moreover, the constant global phase $\theta(\omega_0) = \phi_a(\omega_0) + \phi_b(\omega_0)$ does not appear in the final result, meaning that common phase fluctuations affecting both paths equally have no impact on the measurement.

For our product-state scheme with post-selection, we include the phase noise contributions in the spectral correlation measurement result. Following the same post-selection procedure as in Section 2—selecting the anti-diagonal components $\omega = \omega_0 + \Omega$ and $\omega' = \omega_0 - \Omega$—and using the first-order expansion of the phase noise, we obtain

$$P_{P,anti} = \frac{1}{2} e^{-\frac{\Omega^2}{(2\sigma^2)}} (1 - cos(2\Omega(\tau - \varepsilon_0))), \tag{18}$$

Crucially, the interference term depends on $2\Omega(\tau - \varepsilon_0)$, where $\varepsilon_0$ is exactly the same relative linear phase noise coefficient as in the entanglement-based case. The constant global phase $\theta(\omega_0)$ cancels completely, just as in the entangled HOM case. Applying a Fourier transform to Eq. 18 and retaining only the positive frequency component yields the final temporal interference pattern

$$P_P \propto \frac{1}{2} e^{-\frac{\sigma^2 t^2}{2}} - \frac{1}{4} e^{-\frac{\sigma^2 (t - 2(\tau - \varepsilon_0))^2}{2}}, \tag{19}$$

Comparing Eq. (19) with Eq. (17), we see that both schemes exhibit the same behavior under phase noise: the interference peak is shifted by $\varepsilon_0$ but its width and shape remain unchanged. The linear phase noise effectively introduces a constant time offset that can be calibrated out, while the overall measurement precision is not degraded. This demonstrates that our product-state scheme with post-selection preserves the phase-noise immunity of entanglement-based HOM interferometry. The post-selection operation extracts the frequency-anti-correlated component from the product state, generating an effective interference term that inherits the same robustness against phase perturbations. Fig. 4 presents the simulation results of our scheme under four different noise configurations, defined as $\varepsilon_0 = 0, \theta(\omega_0) = 0$ ; $\varepsilon_0 = -0.2ps, \theta(\omega_0) = 0$ ; $\varepsilon_0 = 0, \theta(\omega_0) = 0.1ps$ and $\varepsilon_0 = -0.2ps, \theta(\omega_0) = 0.1ps$ , corresponding to Figs. 4(a), 4(b), 4(c), and 4(d), respectively. The light source has a central wavelength of 810 nm and a spectral FWHM of 100 nm. The sample has a refractive index of 1.5 and a thickness of $4 \times 10^{-5}\, m$, with no dispersion considered. The value $n = 1.5$ is chosen as a typical refractive index for glass. From Figs. 4(a) and 4(c), it can be seen that when $\varepsilon_0 = 0$, the interference peak position and FWHM remain unchanged for either $\theta(\omega_0) = 0$ or $\theta(\omega_0) = 0.1ps$. Therefore, our scheme exhibits the same robustness to $\theta(\omega_0)$ as entangled HOM interference. Furthermore, the same conclusion can be drawn by comparing Figs. 4(a), 4(b), and 4(d). Under the condition $\theta(\omega_0) = 0$, changing $\varepsilon_0$ from 0 to $-0.2\,ps$ shifts the peak position by 20 μm. However, when changing from the case $\varepsilon_0 = -0.2\,ps, \theta(\omega_0) = 0$ to $\varepsilon_0 = -0.2\,ps, \theta(\omega_0) = 0.1\,ps$, the peak position undergoes no change.

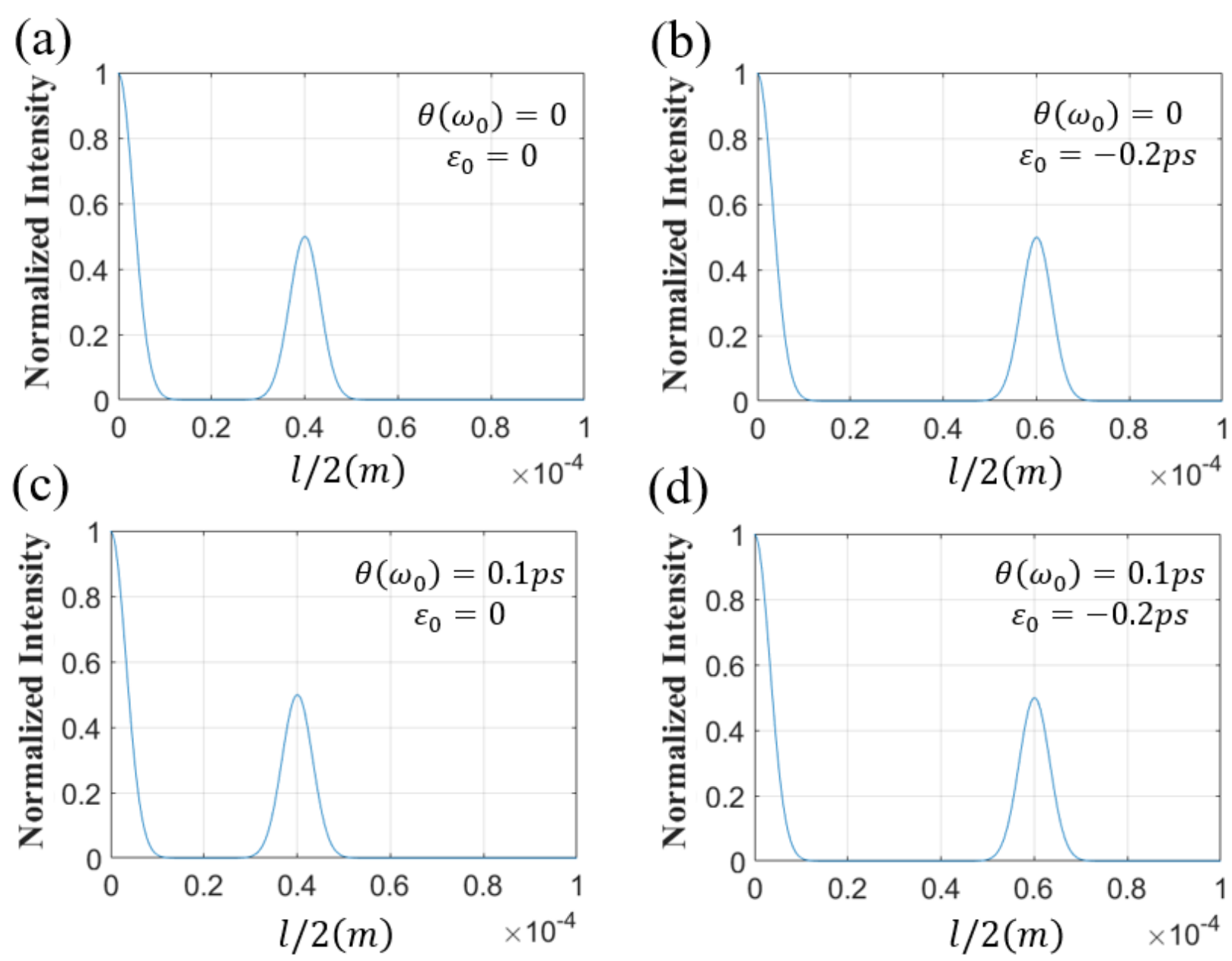

**Fig. 4.** Simulation results of the product state with post selection scheme under different noise conditions. a, $\varepsilon_0 = 0, \theta(\omega_0) = 0$. b, $\varepsilon_0 = -0.2ps, \theta(\omega_0) = 0$. c, $\varepsilon_0 = 0, \theta(\omega_0) = 0.1ps$. d, $\varepsilon_0 = -0.2ps, \theta(\omega_0) = 0.1ps$.

### 3.4 Fisher Information and the Precision Cost of the Product-State Scheme

For quantum sensing, the core performance metric is measurement precision (or equivalently, the signal-to-noise ratio). Conventional HOM interferometers typically employ frequency-entangled states as probes, and their phase estimation precision is fundamentally limited by the standard quantum limit (SQL) [31,57]. For a phase shift introduced only on a single arm, the quantum Cramér–Rao lower bound can be expressed as

$$\Delta^2\tau \geq \frac{1}{F_Q^{ent}} = \frac{1}{4\sigma^2}, \tag{20}$$

where $\sigma$ is the standard deviation (width parameter) of the Gaussian biphoton spectral distribution, defined through $\varphi(\Omega) = exp\left(-\frac{\Omega^2}{4\sigma^2}\right)$. As derived in Supplemental Document E, the quantum Fisher information (QFI) for the entangled-state HOM interferometer is $F_Q^{ent} = 4\sigma^2$. This precision limit originates from the fact that the single-arm phase-shift operator acts on only one subsystem of the entangled state, restricting the information extraction efficiency to the SQL and preventing the nonlocal correlations of entanglement from providing a further advantage [31].

To clarify the precision cost of replacing entanglement with post-selection, we analyze the Fisher information of our proposed scheme. We emphasize that the goal of this section is not to claim an equal precision to the entangled-state HOM interferometer, but to quantify the actual information available under the spectral-correlation measurement and post-selection. For this purpose, we employ the rigorous tool of Fisher information. We first compute the QFI of the input product state under a single-arm phase shift, which represents the ultimate precision limit achievable with an optimal measurement. Following the derivation in Supplemental

Document E, for the frequency product state given in Eq. (3), with each photon's spectral amplitude taking the Gaussian form

$$\phi_p(\omega_2) = exp\left(-\frac{(\omega_2 - \omega_0)^2}{4\sigma^2}\right), \tag{21}$$

which is identical to the single-photon spectral amplitude in the entangled case, the QFI is

$$F_Q^{prod} = 4\sigma^2, \tag{22}$$

This result shows that, under the same spectral width parameter $\sigma$, the entanglement-free product-state probe possesses the same theoretical phase information (in terms of QFI) as the entangled-state probe. This also shows that the measurement precision itself does not originate from entanglement. However, the QFI is only a theoretical upper bound, as it requires performing a joint optimal measurement over the entire system, which is often experimentally impractical.

In practice, our scheme does not implement a globally optimal measurement; instead, it employs spectral correlation measurements followed by post-selection. Therefore, the actual precision of our scheme is determined by the classical Fisher information (CFI) corresponding to this specific measurement strategy. As detailed in Supplemental Document E, after post-selecting the three possible output channels (one photon in each output port, both photons in port $c$, and both photons in port $d$), the conditional classical Fisher information (CFI) per successful post-selection event is

$$F_{cl,post} = 2\sigma^2, \tag{23}$$

This value is exactly half of the QFI of the entangled-state HOM interferometer. Moreover, when the success probability of post-selection is taken into account, the total CFI per input photon pair becomes is

$$F_{actual} = P_{success} \cdot F_{c,post} = P_{success} \cdot 2\sigma^2, \tag{24}$$

where $P_{success}$ denotes the probability that a photon pair satisfies the post-selection condition. The post-selection process inevitably introduces a probabilistic cost. This means that, although the conditional measurement can reach $2\sigma^2$, the effective Fisher information per input pair is reduced by the post-selection success probability.

However, the probabilistic penalty induced by post-selection can be eliminated by extending the measurement from a single anti-diagonal line to all diagonals of the spectral correlation map. In the previous scheme, the processor selects only the central anti-diagonal components satisfying $\omega = \omega_0 + \Omega$ and $\omega' = \omega_0 - \Omega$, which is equivalent to fixing $u = \omega - \omega' = 2\Omega, v = \omega + \omega' = 2\omega_0$. In this case, only a subset of photon pairs—those whose two frequencies are anti-correlated around $\omega_0$ —contributes to the final interference signal, while all other events are discarded. In contrast, the full-diagonal scheme retains the entire spectral correlation measurement and reorganizes it in terms of the diagonal and anti-diagonal variables,

$$u = \omega - \omega', v = \omega + \omega', \tag{25}$$

Instead of selecting a single line $v = 2\omega_0$. The essential difference from the previous scheme is that the post-selection is no longer a binary filtering of events. It becomes a deterministic reweighting and integration over the full spectral correlation plane: every detected photon pair contributes to the Fisher information, and no event is discarded. The total CFI is given by

$$F_{total} = 2\sigma^2, \tag{E21}$$

The detailed calculation progress is presented in Supplemental Document E. This means that the full-diagonal scheme removes the probabilistic overhead of post-selection and recovers the full conditional Fisher information $2\sigma^2$ per input photon pair. Nevertheless, it should be emphasized that even with all diagonals and all three measurement channels, the total CFI remains $2\sigma^2$, which is still half of the QFI of the entangled-state HOM interferometer, $4\sigma^2$.

## 4 Discussion

In this work, we have shown that the metrological advantages of Hong–Ou–Mandel interferometry—dispersion cancellation, phase-noise immunity, peak narrowing, and wide dynamic range—can be reproduced without entanglement. Specifically, we employed frequency product states as the probe and applied post-selection to extract anti-diagonal components from the spectral correlation measurement. The resulting interference pattern is identical to that obtained with entangled biphoton states. This demonstrates that the essential resource for these advantages—dispersion cancellation, phase-noise immunity, peak narrowing, and wide dynamic range—is not entanglement or exchange symmetry, but effective frequency anti-correlation, which can be synthesized from product states via post-selection. Through a rigorous Fisher information analysis, we further show that the measurement precision itself does not originate from entanglement: under the same single-photon bandwidth, the product state and the entangled state possess the same quantum Fisher information, $F_Q = 4\sigma^2$. For our specific scheme, the conditional classical Fisher information of the anti-diagonal scheme is half of the quantum Fisher information of the entangled-state HOM interferometer. The difference arises solely from the measurement strategy, post-selection is not a globally optimal joint measurement. When the post-selection success probability is included, the actual Fisher information per input photon pair is further reduced to $P_{success} \times 2\sigma^2$. Extending the measurement to all diagonals eliminates this success-probability penalty and yields a total classical Fisher information of $F_{total} = 2\sigma^2$, but this is still only half of the entangled-state QFI. The remaining factor of two reflects the fact that the product-state plus spectral-correlation measurement is not a globally optimal joint measurement. Our results provide a rigorous mathematical counterpart to Franson's qualitative argument [41] and bridge the gap between entangled HOM interferometry and its classical emulations.

From a practical perspective, our scheme eliminates the need for entanglement or bosonic exchange symmetry, making HOM-type measurements accessible with simple broadband coherent sources. This may facilitate real-world applications in biomedical imaging, lidar, and industrial inspection, where robustness and experimental simplicity are often more critical than the use of quantum resources

**Data availability.**

The data that underlie the plots within the paper and other findings of this study are available from the corresponding authors on reasonable request.

**Code availability**

The code used to generate simulated data and plots is available from the corresponding authors on reasonable request.

**Competing interests' statement.** The authors declare no competing interests.